\documentclass[pre,twocolumn,floatfix,byrevtex,final,aps]{revtex4}
\usepackage{amsmath,graphicx}
\begin{document}

\title{Spatio-temporal dynamics of coupled array of Murali-Lakshmanan-Chua
circuits}

\author{P. Muruganandam}

\email{anand@cnld.bdu.ac.in}

\affiliation{Centre for Nonlinear Dynamics, Department of Physics,
Bharathidasan University, Tiruchirapalli 620 024, India}

\author{K. Murali}
%\email{e-mail: phymurli@hotmail.com}

\altaffiliation[Present address: ]{Department of Physics, Anna University, Chennai 600 025, India}
\affiliation{Centre for Nonlinear Dynamics, Department of Physics,
Bharathidasan University, Tiruchirapalli 620 024, India}

\author{M. Lakshmanan}

\email{lakshman@cnld.bdu.ac.in}

\affiliation{Centre for Nonlinear Dynamics, Department of Physics,
Bharathidasan University, Tiruchirapalli 620 024, India}

\date{}

\begin{abstract}

The circuit recently proposed by Murali, Lakshmanan and Chua (MLC) is
one of the simplest non-autonomous nonlinear electronic circuits which
shows a variety of dynamical phenomena including various bifurcations,
chaos and so on.  In this paper we study the spatio-temporal dynamics
in one and two dimensional arrays of coupled MLC circuits both in the
absence as well as in the presence of external periodic force. In the
absence of any external force, the propagation phenomena of travelling
wave front and its failure have been observed from numerical
simulations.  We have shown that the propagation of travelling wave
front is due to the loss of stability of the steady states (stationary
front) via subcritical bifurcation coupled with the presence of
necessary basin of attraction of the steady states.  We also study the
effect of weak coupling on the propagation phenomenon in one
dimensional array which results in the blocking of wave front due to
the existence of a stationary front. Further we have observed the
spontaneous formation of hexagonal patterns (with penta-hepta defects)
due to Turing instability in the two dimensional array. We show that a
transition from hexagonal to rhombic structures occur by the influence
of external periodic force.  We also show the transition from hexagons
to rolls and hexagons to breathing (space time periodic oscillations)
motion in the presence of external periodic force.  We further analyse
the spatio-temporal chaotic dynamics of the coupled MLC circuits (in
one dimension) under the influence of external periodic forcing.  Here
we note that the dynamics is critically dependent on the system size.
Above a threshold size, a suppression of spatio-temporal chaos occurs,
leading to a space-time regular (periodic) pattern eventhough the
single MLC circuit itself shows a chaotic behaviour. Below this
critical size, however, a synchronization of spatiotemporal chaos is
observed.  

\end{abstract} 

\maketitle

\section{Introduction}
\label{sec1}

Coupled nonlinear oscillators in the form of arrays arise in many
branches of science. These coupled arrays are often used to model very
many spatially extended dynamical systems explaining reaction-diffusion
processes.  One can visualize such an array as an assembly of a number
of subsystems coupled to their neighbours, or as coupled nonlinear
networks (CNNs) [Chua \& Yang, 1988; Chua, 1995].  Theoretically one
can consider a system of coupled ordinary differential equations (odes)
to represent a macroscopic system, in which each of the odes
corresponds to the evolution of the individual subsystem.  Specific
examples are the coupled array of anharmonic oscillators [Mar\'{\i}n \&
Aubry, 1996], Josephson junctions [Watanabe {\it et al.}, 1995],
continuously stirred tank reactors (CSTR) exhibiting travelling waves
[Dolnik \& Marek, 1988; Laplante \& Erneux, 1992], propagation of nerve
impulses (action potential) along the neuronal axon[Scott, 1975], the
propagation of cardiac action potential in the cardiac tissues [Allesie
{\it et al.}, 1978] and so on.  For the past few years several
investigations have been carried out to understand the spatio-temporal
behaviours of these coupled nonlinear oscillators and systems. The
studies on these systems include the travelling wave phenomena, Turing
patterns, spatio-temporal chaos and synchronization [Feingold {\it et
al.}, 1988; Mu\~{n}uzuri {\it et al.}, 1993; P\'{e}rez-Mu\~{n}uzuri
{\it et al.}, 1993; Sobrino {\it et al.}, 1993; P\'{e}rez-Mu\~{n}uzuri
{\it et al.}, 1994, P\'{e}rez-Mu\~{n}uzuri {\it et al.}, 1995; Kocarev
\& Parlitz, 1996].

One of the interesting behaviours of these coupled oscillators is that
they can exhibit a failure mechanism for travelling waves which cannot
be observed in the homogeneous continuous models [Keener, 1987].
Similar failure phenomena in the nerve impulse propagation have been
observed in biological experiments [Balke {\it et al.}, 1988; Cole {\it
et al.}, 1988].

Recently various spatio-temporal patterns have been found in discrete
realizations of the reaction-diffusion models by means of arrays of
coupled nonlinear electronic circuits [P\'erez-Mu\~{n}zuri {\it et al.},
1993].  For example P\'erez-Mu\~{n}uzuri and coworkers studied the
various spatio-temporal patterns in a model of discretely coupled
Chua's circuits. This system is a {\it three} variable model and
coupled to its neighbours by means of a linear resistor. In this paper
we consider similar arrays of coupled nonlinear electronic circuits
consisting of Murali-Lakshmanan-Chua (MLC) circuits as a basic unit,
which is basically a {\it two} variable model. The dynamics of the
driven MLC circuit including bifurcations, chaos, controlling of chaos
and synchronization has been studied by Murali {\it et al.} [1994a;
1994b; 1995; Lakshmanan \& Murali, 1996]. Of particular interest among
coupled arrays is the study of {\it diffusively coupled driven systems}
as they represent diverse topics like Faraday instability
[Lioubashevski {\it et al.}, 1996], granular hydrodynamics [Umbanhawar
{\it et al.}, 1996; Kudrolli {\it et al.}, 1997], self organized
criticality [Bak \& Chen, 1991] and so on.  Identification of localized
structures in these phenomena has been receiving considerable attention
very recently [Fineberg, 1996]. Under these circumstances, study of
models such as the one considered in this paper can throw much light on
the underlying phenomenon in diffusively coupled driven systems.

In the following by considering one and two dimensional arrays of
coupled MLC circuits, we have studied the various spatio-temporal
patterns in the presence as well as in the absence of external force.
We present a brief description on the one and two dimensional arrays of
MLC circuits in Sec. \ref{sec2}. Then, in Sec. \ref{sec3}, in the
absence of periodic external force, we study the propagation phenomenon
of travelling wave front and its failure below a critical diffusive
coupling coefficient. We show that the propagation phenomenon occurs
due to a loss of stability of the steady states via subcritical
bifurcation coupled with the presence of necessary basin of attraction
of the steady states for appropriate diffusive coupling coefficient.
We also study the effect of weak coupling on the propagation phenomenon
which causes a blocking in the propagation due to the presence of a
stationary front at the weakly coupled cell.  In addition we also study
the onset of Turing instability leading to the spontaneous formation of
hexagonal patterns in the absence of any external force.

Further as mentioned earlier it is of great physical interest to study
the dynamics of the coupled oscillators when individual oscillators are
driven by external forces.  In Sec. \ref{sec4}, we study the
spatio-temporal patterns in the presence of periodic external force 
and investigate the effect of it on the propagation phenomenon and
Turing patterns. Depending upon the choice of control parameters,
a transition from hexagons to regular rhombic structures, hexagons
to rolls and then to breathing oscillations from hexagons are observed.
The presence of external force with sufficient strength removes the
penta-hepta defect pair originally present in the spontaneously formed
hexagonal patterns leading to the formation of regular rhombic structures.
The inclusion of external periodic force can also induce a
transition from hexagons to rolls provided there are domains of small
roll structures in the absence of force. We further show that in the
region of Hopf-Turing instability, the inclusion of external periodic
force with sufficiently small amplitude induces a type of breathing
oscillations though the system shows a regular hexagonal pattern in the
absence of any external force.

Finally, we also study the spatio-temporal chaotic dynamics of the one
dimensional array of MLC circuits in Sec. \ref{sec5} when individual
oscillators oscillate chaotically.  In this case, the emergence of
spatio-temporal patterns depends on the system size. For larger size,
above a critical number of cells, we observe a controlled space-time
regular pattern eventhough the single MLC circuit itself oscillates
chaotically. However, synchronization occurs for a smaller system size,
below the threshold limit. Finally, in Sec. \ref{sec6} we give our
conclusions.

%%%%%%%%%%%%%%%%%%%%%%%%%%%%%%%%%%%%%%%%%%%%%%%%%%%%%%%%%%%%%%%%%%%%%
\section{Arrays of Murali-Lakshmanan-Chua circuits}
\label{sec2}

As mentioned in the introduction the circuit proposed by Murali,
Lakshmanan and Chua (MLC) is one of the simplest second order
dissipative nonautonomous circuit having a single nonlinear element
[Murali {\it et al}., 1994a].  Brief details of its dynamics are given
in the Appendix A for convenience. Here we will consider one and two
dimensional arrays of such MLC circuits, where the intercell couplings
are effected by linear resistors.

\subsection{One-dimensional array}
\begin{figure}[!ht]
\includegraphics[width=\linewidth]{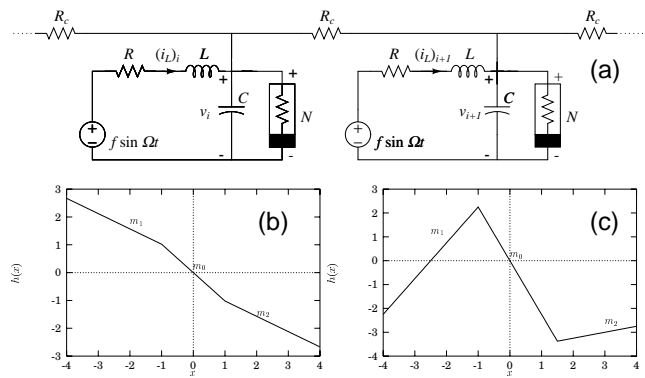}
\caption{
(a) Circuit realization of an one dimensional array of coupled MLC circuits.
(b) Graph of the function $h(x)$ Vs. $x$ for the parametric choice
$\{m_0, m_1, m_2, \epsilon \} = $ $\{-1.02, -0.55, -0.55, 0\}$ in Eq. (2).
(c) Graph showing $h(x)$ Vs. $x$ for the parametric choice $\{m_0, m_1, m_2, 
\epsilon \} =\{-2.25, 1.5, $ $0.25, 0\}$ in Eq. (2). 
}
\label{fig1}
\end{figure}

Fig. \ref{fig1}(a) shows a schematic representation of an one
dimensional  chain of resistively coupled MLC circuits.  Extending the
analysis of single MLC circuit as given in the Appendix A, the dynamics
of the one dimensional chain can be easily shown to be governed by the
following system of equations, in terms of the rescaled variables (see
Appendix A),
\begin{subequations}
\label{eqn1all}
\begin{eqnarray}
\dot x_{i} & = & y_{i}-h(x_{i})+D(x_{i+1}+x_{i-1}-2x_{i}), \\
\dot y_{i} & = & -\sigma y_{i}-\beta x_{i}+F\sin \omega t, 
\,\,\,\,\, i=1,2,\cdots,N,
\end{eqnarray}
\end{subequations}
where $D$ is the diffusion coefficient, $N$ is the chain length and
$h(x)$ is a three segment piecewise linear function representing the
current voltage characteristic of the Chua's diode,
\begin{equation}
\label{eqn2}
h(x)=\left\{
\begin{matrix}
\epsilon+m_2x+(m_0-m_1), & x\geq x_2         \cr
\epsilon+m_0x,           & x_1\leq x\leq x_2 \cr
\epsilon+m_1x-(m_0-m_1), & x\leq x_1.
\end{matrix}
\right.
\end{equation}
In (\ref{eqn2}) $m_0$, $m_1$ and $m_2$ are the three slopes. Depending
on the choice of $m_0$, $m_1$ and $m_2$ one can fix the characteristic
curve of the Chua's diode.  Here $\epsilon $ corresponds to the dc
offset.  Of special importance for the present study are the two forms
of $h(x)$ shown in Figs. \ref{fig1}(b) and \ref{fig1}(c). Fig.
\ref{fig1}(b) corresponds to the choice of the parameters $\{m_0, m_1,
m_2, \epsilon\} = \{-1.02, -0.55, -0.55, 0\}$ while Fig. \ref{fig1}(c)
corresponds to the parameters $\{m_0, m_1, m_2, \epsilon\} =\{-2.25,
1.5, 0.25, 0\}$.  As noted in the Appendix A, the significance of the
characteristic curve in Fig. \ref{fig1}(b) is that it admits
interesting bifurcations and chaos, while Fig. \ref{fig1}(c)
corresponds to the possibility of bistable states with asymmetric
nature which is a prerequisite for observing wave front propagation and
its failure.  In the following studies we will consider a chain of
$N=100$ MLC circuits.

\subsection{Two-dimensional array} As in the case of one dimensional
array one can also consider a two dimensional array with each cell in
the array being coupled to four of its nearest neighbours with linear
resistors. The model equation can be now written in dimensionless form
as
\begin{widetext}
\begin{subequations}
\label{eqn3all}
\begin{eqnarray}
&& \dot x_{i,j} = y_{i,j}-h(x_{i,j})+D_1(x_{i+1,j}+x_{i-1,j}+x_{i,j+1}
+x_{i,j-1}-4x_{i,j})
\equiv  f(x_{i,j},y_{i,j}) , \\
&& \dot y_{i,j} = -\sigma
y_{i,j}-\beta x_{i,j}+D_2 (y_{i+1,j}+y_{i-1,j}+y_{i,j+1}
+y_{i,j-1}-4y_{i,j})+F\sin \omega t
\equiv  g(x_{i,j},y_{i,j}),\;\;\;
 i,j =  1,2,\cdots, N.\nonumber 
\end{eqnarray}
\end{subequations}
\end{widetext}
This two dimensional array has $N\times N$ cells arranged in a
square lattice.  In our numerical study we will again take $N=100$.

      In the following sections we present some of the interesting
dynamical features of the array of MLC circuits such as propagation of
wave front and its failure, effect of weak coupling in the propagation,
Turing patterns, effect of external periodic forcing in the Turing
patterns and spatio-temporal chaotic dynamics. We have used zero flux
boundary conditions for the study of propagation phenomenon and Turing
patterns and periodic boundary conditions for the study of
spatio-temporal chaos.

%%%%%%%%%%%%%%%%%%%%%%%%%%%%%%%%%%%%%%%%%%%%%%%%%%%%%%%%%%%%%%%%%%%%%
\section{Spatio-temporal patterns in the absence of external force:
Travelling wave phenomenon and Turing patterns}
\label{sec3}

Transport processes in living tissues, chemical and physiological
systems have been found to be associated with special types of waves
called travelling waves.  Earlier, continuous models were created to
describe the wave propagation phenomenon in these systems, but these
failed to cover all the important aspects of travelling wave
behaviour.  One of the most important of them is the so called
travelling wave propagation failure, occurring at weak coupling between
cells. It has been proved by Keener [1987] that propagation failure
cannot be observed in a continuous, one variable, homogeneous
reaction-diffusion system. Therefore to study this kind of phenomenon,
one has to use discrete models. Recently travelling wave phenomenon and
its failure have been studied in arrays of discretely coupled Chua's
circuits [P\'{e}rez-Mu\~{n}uzuri {\it et al.}, 1993].  Wave propagation
and its failure have also been observed even in one variable models
like discrete Nagumo equation [Leenaerts, 1997].  In this section we
carry out a study of such wave propagation phenomenon in one and two
dimensional arrays of Murali-Lakshmanan-Chua's (MLC) circuit
oscillators, without forcing, and investigate the mechanism by which
such a failure occurs.

\subsection{Propagation phenomenon and its failure
in one-dimensional array}
\label{sec3a}

In order to observe wave front propagation in the one dimensional array
of MLC circuits, we numerically integrated the Eqs. (1) using fourth
order Runge-Kutta method.  In this analysis we fix the parameters at
$\{\beta, \sigma, m_0, m_1, m_2, \epsilon,$ $F \}= \{1.0, 1.0, -2.25,
1.5, .25, 0, 0\}$ so that the system admits bistability and this choice
leads to the asymmetry characteristic curve for the Chua's diode as
shown in Fig.  \ref{fig1}(c). The existence of bistable state in the
asymmetric case is necessary to observe a wave front.  Zero flux
boundary conditions were used in the numerical computations, which in
this context mean setting $x_0=x_1$ and $x_{N+1}=x_N$ at each
integration step; similar choice has been made for the variable $y$
also. To start with, we will study in this section the force-free case,
$F=0$, and extend our studies to the forcing case $(F\neq 0)$ in the
next section.

The choice of the values of the parameters guarantees the existence of
two stable equilibrium points $P^+_i =\{\sigma (m_1 - m_0 - \epsilon)/
(\beta + m_2 \sigma )$, $\beta (m_0 - m_1 - \epsilon)/ (\beta + m_2
\sigma )\}$ and $P^-_i = \{\sigma (m_1 - m_0 - \epsilon)/(\beta +
m_1\sigma )$, $ \beta (m_0 - m_1 - \epsilon)/(\beta + m_1\sigma )\}$ for
each cell.  In the particular case, corresponding to  the above
parametric choice, each cell in the array has three equilibrium points
at $P^+_i=(3.0,-3.0)$, $P^-_i=(-1.5,1.5)$ and $P^0_i=(0,0)$. Out of
these three equilibrium points, $P^+_i$ and $P^-_i$ are stable while
$P^0_i$ is unstable.  Due to the asymmetry in the function $h(x)$ for
the present parametric choices, the basin of attraction of the point
$P^+_i$ is much larger than that of $P^-_i$ and it is harder to steer a
trajectory back into the basin $P^-_i$ once it is in the basin of
$P^+_i$.

Now we choose an initial condition such that the first few cells in the
array are excited to the $P_i^+$ state (having a large basin of
attraction compared to that of $P_i^-$) and the rest are set to $P_i^-$
state. In other words a wave front in the array is obtained by means of
the two stable steady states. On actual numerical integration of
Eqs.(\ref{eqn1all}) with $N=100$ and with the diffusion coefficient
\begin{figure}[!ht]
\includegraphics[width=\linewidth]{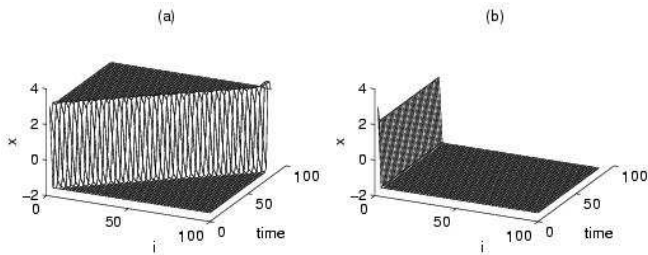}
\caption{ Space-time plot showing (a) propagation of wave fronts in one
dimensional array (100 cells) of MLC circuits for $D=2.0$ and (b)
propagation failure for $D=0.4$.  }
\label{fig2} 
\end{figure}
chosen at a higher value, $D=2.0$, a motion of the wave front towards
right (see Fig. \ref{fig2}(a)) is observed, that is a travelling wave
front is found.  After about 80 time units the wave front reaches the
$100^{\mbox{th}}$ cell so that all the cells are now settled at the
more stable state $(P_i^+)$ as demonstrated in Fig. \ref{fig2}(a). When
the value of $D$ is decreased in steps and the analysis is repeated,
the phenomenon of travelling wavefronts continues to be present.

However, below a critical value of the diffusion coefficient
$(D=D_c)$ a failure in the propagation has been observed, which in the
present case turns out to be $D=D_c=0.4$. This means that the initiated
wavefront is unable to move as time progresses and Fig. \ref{fig2}(b)
shows the propagation failure for $D=0.4$.

\subsection{Propagation failure mechanism: A case study of
5 coupled oscillators}
\label{sec3b}

To understand the failure mechanism, we have carried out a detailed
case study of an array of five coupled MLC circuits, represented by the
following set of equations. We have verified our conclusions for the
case of 3, 4, 6 and 7 oscillators also.  The analysis can be in
principle extended to arbitrary but finite number of oscillators. From
Eq.  (\ref{eqn1all}), the present array can be represented by the
following set of ten coupled first order odes:
\begin{eqnarray}
\label{eqsta}
\dot x_1 & = & y_1-h(x_1)+D(x_2-x_1), \nonumber \\
\dot x_2 & = & y_2-h(x_2)+D(x_1-2x_2+x_3), \nonumber \\
\dot x_3 & = & y_2-h(x_3)+D(x_2-2x_3+x_4), \nonumber \\
\dot x_4 & = & y_4-h(x_4)+D(x_3-2x_4+x_5), \nonumber \\
\dot x_5 & = & y_5-h(x_5)+D(x_4-x_5), \nonumber \\
\dot y_j & = & -\sigma y_j-\beta x_j,\,\,\,\,\,\,j=1,2\ldots,5,
\end{eqnarray}
where $h(x)$ is as given in Eq.  (\ref{eqn2}) and the parameters are
fixed at $\beta=1$, $\sigma=1$, $m_0=-2.25$, $m_1=1.5$, $m_2=0.25$ and
$\epsilon=0$.

\subsubsection{Numerical analysis}
\label{secnum}

      By carrying out a numerical analysis of the system (\ref{eqsta})
with zero flux boundary conditions, we identify the following main
results.
\paragraph{}
\label{casea}
We solve the above system of equations (\ref{eqsta}) for the chosen
initial condition $\{x_1(0),$ $x_2(0),$ $x_3(0),$ $x_4(0),$ $x_5(0) \}$
$=$ $\{3.0,$ $3.0,$ $-1.5,$ $-1.5,$ $-1.5\}$ and $y_i(0)=-x_i(0)$,
$i=1,2\ldots 5,$ for different values of the diffusion coefficient $D$
in the region $D\in (0, 1.0)$. Note that the values $x_1(0)=x_2(0)=3.0$
and $x_3(0)=x_4(0)=x_5(0)=-1.5$ correspond to the stable equilibrium
points $P_i^+$ and $P_i^-$, respectively, of the uncoupled MLC
oscillators.  When $D$ is decreased downwards from $1.0$ we find that
in the region $D>0.4$ propagation of wavefront occurs as in Fig.
\ref{fig2}(a) (of the $N=100$ cells system). At the critical value
$D=0.4$, the wavefront fails to propagate and asymptotically ends up in
a stationary wavefront which is nearer to the initial condition (as in
Fig. \ref{fig2}(b)). The wavefront failure phenomenon persists for all
values of $0 \le D \le 0.4$.

\paragraph{}
\label{caseb}

On the other hand, if we choose the initial condition as $\{x_1(0),$
$x_2(0),$ $x_3(0),$ $x_4(0),$ $x_5(0) \}$ $=$ $\{2.7990,$ $2.1709,$
$1.7803,$ $-1.4433,$ $-1.4923\}$ and $y_i(0)=-x_i(0)$, $i=1,2,..5,$
then the critical value $D$, where wavefront propagation failure
occurs, turns out to be at a slightly higher value $D=D_c=0.4703$.

\subsubsection{Linear stability analysis}
\label{seclin}

In order to understand the failure mechanism of wavefront propagation
we consider all the possible steady states (equilibrium states)
associated with Eqs.(\ref{eqsta}) for nonzero values of $D$. It is easy
to check that Eq. (\ref{eqsta}) possesses a maximum of $3^5=243$ steady
states.  First we consider the three trivial steady states which can be
easily obtained by considering $x_1=x_2=x_3=x_4=x_5$ and
$y_1=y_2=y_3=y_4=y_5$, namely, $X_p^+$ $=$ $(3.0, -3.0, 3.0, -3.0, 3.0,
-3.0, 3.0, -3.0, 3.0, -3.0)$, $X_p^0$ $=$ $(0, 0, 0,0,0,0,0,0,0,0)$ and
$X_p^-$ $=$ $(-1.5, 1.5, -1.5, 1.5, -1.5, 1.5, -1.5, 1.5, -1.5, 1.5)$.
(Here the suffix $p$ indicates that the steady states are the same as
that of the $D=0$ case, namely $P_i^+$, $P_i^0$ and $P_i^-$). From a
linear stability analysis (details are given in Appendix B) we can
easily conclude that the steady states $X_p^+$ and $X_p^-$ are stable
while $X_p^0$ is unstable irrespective of the value of the diffusion
coefficient $D$. Then any initial condition in the neighbourhood of
$X_p^\pm$ will evolve into $X_p^\pm$ asymptotically. Also one has to note
that, due to asymmetry in the function $h(x)$, the basin of $X_p^+$ is much
larger than that of $X_p^-$.

Since we are interested here to understand the propagation failure
mechanism, we concentrate on those steady states which possess
stationary wavefront structures {\em near} to the two types of chosen
initial conditions in the numerical analysis discussed in Sec.
\ref{secnum}. Out of the 243 available steady states, we select a
subset of them, $X_s=\{X_s^+, X_s^0, X_s^-\}$, which is nearer to the
numerically chosen initial state. These three steady states are defined
in Eqs.(\ref{steady}) of Appendix B. They are identified such that the
components $x_{1s}$ and $x_{2s}$ are in the right extreme segment of
the characteristic curve, Fig. \ref{fig1}(c), while $x_{4s}$ and
$x_{5s}$ are in the left extreme segment. On the other hand $x_{3s}$ is
allowed to be in any one of the three segments.  The reason for such a
choice is that we are looking for the formation of a wavefront which is
closer to the chosen initial conditions in Sec. \ref{secnum}.  The
corresponding forms of $h(x_i)$ as given in Eq. (\ref{eqn2}) are then
introduced in Eqs.(\ref{eqsta}) to find the allowed steady states. Then
$X_s^+$ corresponds to the case in which $x_{3s}$ as well as $x_{1s}$
and  $x_{2s}$ are in the right most segment while the rest are in the
left most segment and $X_s^0$ and $X_s^-$ correspond to the cases in
which $x_{3s}$ is, respectively, in the middle and the left extreme
segments of the characteristic curve while $x_{1s}$, $x_{2s}$ and
$x_{4s}$, $x_{5s}$ are chosen as in $X_s^+$.  These steady states may
be explicitly obtained as a function of the diffusion coefficient $D$,
whose forms are given in Appendix B.

We now consider the linear stability of these steady states $X_s^+$,
$X_s^0$ and $X_s^-$. The corresponding stability (Jacobian) matrices
are obtained from the linearized version of Eq. (\ref{eqsta}), whose
form is given in Eq. (\ref{jacob}). The eigenvalues are evaluated by
numerical diagonalization of (\ref{jacob}) as a function $D$ and the
stability property  analysed. It has been found that out of these three
steady states both $X_s^+$ and $X_s^-$ are stable for $D<0.4703$ in
which all of the eigenvalues of (\ref{jacob}) are having negative real
parts, while $X_s^0$ is unstable for all values of $D>0$.  However, the
real part of atleast one of the eigenvalues associated with each of
$X_s^+$ and $X_s^-$ changes its sign to positive value at $D=0.4703$
and thereby both the stationary fronts, $X_s^+$ as well as $X_s^-$,
also lose their linear stability at this critical value. Thus for
$D>0.4703$, all the three steady states $X_s^+$, $X_s^0$ and $X_s^-$
are linearly unstable.

First let us consider the second of the initial conditions chosen in
Sec. \ref{secnum}. At and above the critical value $D=0.4703$, all the
three steady states are unstable and so the system transits to the
other nearby steady states. These are also found to be unstable by a
similar analysis, ultimately ending up in the only nearby available
stable state which is $X_p^+$ $=$ $((3.0,-3.0),$ $(3.0,-3.0),$
$(3.0,-3.0),$ $(3.0,-3.0),$ $ (3.0,-3.0))$, thereby explaining the
propagation of the wavefront starting from the initial condition $X(0)$
$=$ $\{(2.7990,$ $-2.7990),$ $(2.1709,-2.1709),$ $(1.7803,-1.7803),$
$(-1.4433,1.4433),$ $(-1.4923,1.4923)\}$ (which is the second of the
numerical results given in Sec. \ref{secnum}).  However for $D <
0.4703$, since $X_s^+$ is stable and is also quite closer to the
initial state, the system settles down in this state itself and no
propagation occurs, thereby explaining the phenomenon of propagation
failure as a subcritical bifurcation when the diffusion coefficient $D$
is increased from $0$ upwards. Note that eventhough $X_s^-$ is also
stable, it is far away from the initial state and so no transition to
this state will occur. A table of these steady states $X_s^+$ and
$X_s^-$ as a function of $D$ (Table \ref{table2}) is given confirming
that $X_s^+$ is always nearer to the initial state compared to
$X_s^-$.

Now let us consider the first of the initial conditions of our
numerical analysis (see Sec. \ref{secnum}), where wavefront propagation
occurs at $D=0.4$ itself and not at $D=0.4703$, eventhough the steady
states $X_s^+$ and $X_s^-$ are still linearly stable in the region
$0.4\le D < 0.4703$.  In order to understand this aspect, we start
analysing the nature of the basin of attraction associated with the
steady state $X_s^-$ for different $D$ values numerically (Note that
for the present set of initial conditions $X_s^-$ is nearer to it than
$X_s^+$). We find that the basin of $X_s^-$ shrinks as $D$ is increased
from $0$ upwards and vanishes completely at $D=0.4703$. We also find
that the chosen initial condition, $X(0)$ $=$ $((3.0,-3.0),$
$(3.0,-3.0),$ $(-1.5,1.5),$ $(-1.5,1.5),$ $(-1.5,1.5)$ falls fully
within the basin of $X_s^-$ as long as $D<0.4$.  However due to the
\begin{figure}[!ht]
\includegraphics[width=\linewidth]{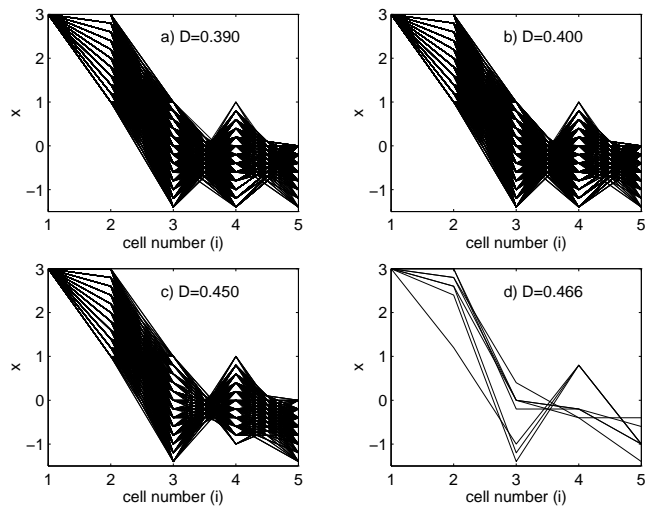}
\caption{Figure showing partial basin for the steady state $X_s^-$ for
various values of the diffusive coupling coefficient. In the figure
each line corresponds to an initial condition. The figure is drawn by
fixing the initial state of the first oscillator at $x_1(0)=3.0$,
$y_1(0)=-3.0$ and for clarity decreasing the rest downwards in steps of
0.2.
}
\label{figbas}
\end{figure}
shrinking nature of the basin as $D$ increases (as given in Figs.
(\ref{figbas})), a part of the initial state falls outside the basin
for $D>0.4$ (Figs. \ref{figbas}(b)-(d)). As a result the chosen initial
condition does not end up in the stationary front $X_s^-$ for
$0.4<D<0.4703$ indicating a global instability (eventhough the state
is linearly stable). Therefore propagation starts to occur as soon as
$D>0.4$, while it fails for $D\le 0.4$ and it finally ends in the
stable state $X_p^+$ described in the previous case.  However, if an
initial condition which happens to fall completely within the basin of
attraction of $X_s^\pm$, the propagation starts to occur only for values
of $D>0.4703$. In fact this is what happens for the second chosen
initial condition ($X(0)$ $=$ $\{(2.7990,-2.7990),$ $(2.1709,-2.1709),$
$(1.7803,-1.7803),$ $(-1.4433,1.4433),$ $(-1.4923,1.4923)\}$, as
explained above (Note that the basin structure shown in Fig.
\ref{figbas} does not cover this case).

\begin{figure}[!ht]
\includegraphics[width=\linewidth]{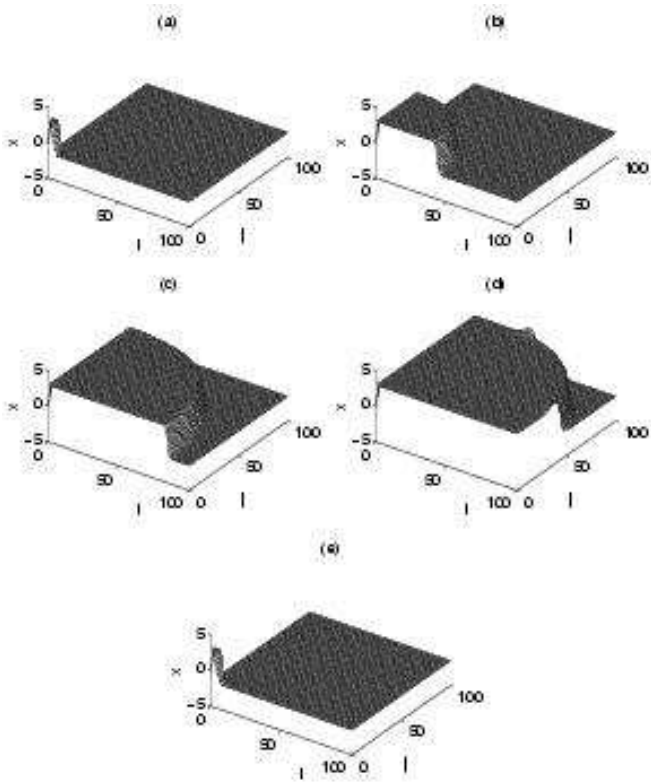}
\caption{
Three dimensional plot showing propagation of two dimensional
wave fronts ($100\times 100)$ at various time units (a) $T=0$,
(b) $T=50$, (c) $T=100$ and (d) $T=125$ for $D_1=2.0$ and $D_2=0$ and 
(e) the same plot for propagation failure case $D_1=0.4$ and $D_2=0$ at 
$T=1000\,\, t.u$.
}
\label{fig3}
\end{figure}

In conclusion, the propagation failure of wave fronts in arrays of
coupled nonlinear oscillators is essentially due to the existence of
stationary fronts (stable steady states) near to the initial states for
low diffusion coefficients, below a certain critical value, which lose
stability via subcritical bifurcation. This coupled with the existence
of necessary basin of attraction (for a global stability) of the steady
state gives rise to the specific critical value of $D=D_c$ above which
propagation begins for a given initial condition.

\subsection{Propagation phenomenon in two-dimensional array}
\label{sec3c}

For studying the propagation phenomenon in two dimensional array of
coupled MLC circuits, we have considered the case in which $F=0$,
$D_1\neq 0$ and $D_2=0$. The propagation of wave front and its failure
have been observed as in the case of one dimensional array.  A two
dimensional array of $100\times 100$ coupled MLC circuits has been
considered for the numerical simulation with initial conditions chosen
such that the first few cells in the array are excited to the $P^+_i$
state and the other cells are set at the $P^-_i$ state.  Figs.
\ref{fig3}(a)-\ref{fig3}(d) show the propagation of the wave front in
the two dimensional array for $D_1=2.0$ and $D_2=0$ at various time
units. Again, below a certain critical diffusion coefficient, $D_1=0.4$,
propagation failure occurs.  One such case is illustrated in Fig.
\ref{fig3}(e) for $D_1=0.4$ and $D_2=0$.  The phenomenon can again be
explained through a linear stability analysis as in the case of the one
dimensional array.

\subsection{Effect of weak coupling}
\label{sec3d}

In the above, the investigation has been made by considering the system
as an ideal one (as far as the circuit parameters are concerned).  But
from a practical point of view, there are defects in the coupling
parameters which may result in a weak coupling  at any of the cell in
the array. In the following we study the effect of such weak coupling
on the propagation of wave front.

Let us consider a weak coupling at the $k^{\mbox{th}}$ cell. By this we
mean that the $k^{\mbox{th}}$ cell in the array is coupled to its
nearest neighbour $(k+1)^{\mbox{th}}$ and $(k-1)^{\mbox{th}}$ cells by
resistors with slightly higher values than that of the others.  We have
studied the effect of this defect on the propagation of wave fronts.
Numerical simulations have been carried out by considering an one
dimensional array with 100 cells, where the initial conditions are
chosen as in the case of propagation phenomenon in regular one
dimensional arrays (see Sec. \ref{sec3a}).  From the numerical
simulation results, we find that there is an abrupt stop in the
propagation when the wave front reaches the weakly coupled cell. This
happens when the coupling coefficient on either side of the
$k^{\mbox{th}}$ cell in the array has a value even above the critical
value $(D=0.4)$ for propagation failure discussed in Sec. \ref{sec3a}.
\begin{figure}[!ht]
\includegraphics[width=\linewidth]{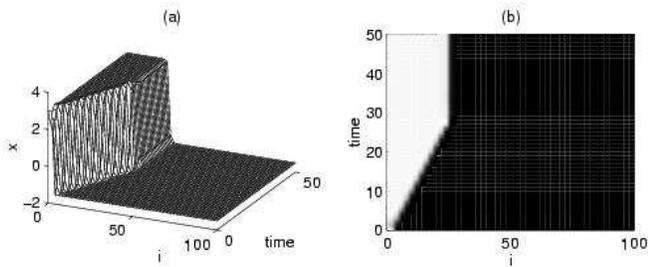}
\caption{
Figure showing the effect of weak coupling at the 25$^{\mbox{th}}$ cell
in the one dimensional array (Eq. (3)). (a) three dimensional
space-time plot and  (b) density plot.
}
\label{fig4}
\end{figure}
Fig. \ref{fig4} shows a blocking in the propagation of the wave front
when the coupling coefficients on either side of the $k^{\mbox{th}}$
cell $(k=25)$, which we call as $D_k$, is set to $0.47173$ with the
rest of the coupling coefficients set to $1$ $(D=1)$.  We observe that
the actual blocking occurs when the wave front reaches the
$k^{\mbox{th}}$ cell. We can say this is a kind of failure in the
propagation (because the wavefront will never reach the last cell (that
is the $100^{\mbox{th}}$ cell in the array) by means of blocking.

As far as the initial conditions are concerned, since the system is in
the propagation region $(D=1.0)$ any choice of initial condition of
wavefront structure will lead to uniform propagation upto the weakly
coupled cell. Now our motivation is to find the property of the above
uniform wavefront when it reaches the weakly coupled cell.

One can analyse the above phenomenon by means of a linear stability
analysis of the steady states as in the case of propagation phenomenon
in Sec. \ref{sec3b}. We consider again as an example the system of five
coupled MLC circuits as in Eqs.(\ref{eqsta}) in which the middle cell
($x_3$) is now taken as an example to be coupled weakly to its
neighbours, respectively $x_2$ and $x_4$. This set up can be
represented by the following set of ten coupled odes:
\begin{eqnarray}
\label{eqweak}
\dot x_1 & = & y_1-h(x_1)+D(x_2-x_1), \nonumber \\
\dot x_2 & = & y_2-h(x_2)+D(x_1-x_2)+D_k(x_3-x_2), \nonumber \\
\dot x_3 & = & y_2-h(x_3)+D_k(x_2-2x_3+x_4), \nonumber \\
\dot x_4 & = & y_4-h(x_4)+D_k(x_3-x_4)+D(x_5-x_4), \nonumber \\
\dot x_5 & = & y_5-h(x_5)+D(x_4-x_5), \nonumber \\
\dot y_j & = & -\sigma y_j-\beta x_j,\,\,\,\,\,\,j=1,2\ldots,5,
\end{eqnarray}
Here $D_k$ corresponds to the coupling coefficient of the weakly coupled
cell and other parameters are the same as in (\ref{eqsta}). As in Sec.
\ref{sec3b}, we calculated the steady states associated with
(\ref{eqweak}) and analysed the linear stability properties.

Let us now consider the three steady states of the system
(\ref{eqweak}), namely, $X_s=\{X_s^+, X_s^0, X_s^-\}$ which possess
wavefront structures near the chosen initial conditions. These steady
states can be found as discussed in Sec. \ref{sec3b}. By fixing
$D=1.0$, we analyse the stability of $X_s$ as a function of $D_k$. We
find that, the steady states, $X_s^\pm$ are linearly stable for values
of $D_k$, $0\le D_k\le 0.4717$ and become unstable for $D_k>0.4717$ via
subcritical bifurcation. However $X_s^0$ state continues to be unstable
irrespective of the value of $D_k$.

At this point, one may think of an array with large number of MLC
circuits. Since all the cells, except the $k^{\mbox{th}}$ cell, have
diffusion coefficient $D=1.0$ which is in the propagation region, the
appropriately chosen initial condition will evolve as a propagating
front upto the weakly coupled cell. When this wave front reaches the
weakly coupled cell, the existence of the stationary front $X_s^-$ for
the array due to the presence of the weakly coupled cell will block the
propagating front and hence it will end up in the stationary front
$X_s^-$ asymptotically for $D_k\le 0.4717$. For $D_k > 0.4717$, the
stationary state $X_s^-$ is no longer stable and hence the propagation
occurs without any blocking. Thus  the blocking of the wave front can
also be attributed to the existence of a stable stationary front when
weak coupling is present in the system.

\subsection{Turing patterns}
\label{sec3e}

Another interesting dynamical phenomenon in the coupled arrays is the
formation of Turing patterns. These patterns are observed in many
reaction diffusion systems when a homogeneous steady state which is
stable due to small spatial perturbations in the absence of diffusion
becomes unstable in the presence of diffusion[Turing, 1952].  To be
specific, the Turing patterns can be observed in a two variable
reaction-diffusion system when one of the variables diffuses faster
than the other and undergo Turing bifurcation, that is, diffusion
driven instability[Turing, 1952; Murray, 1989].

Treating the coupled array of MLC circuits as a discrete version of a
reaction-diffusion system, one can as well observe the Turing patterns
in this model also.  For this purpose, one has to study the linear
stability of system (3) near the steady state.  In continuous systems,
the linear stability analysis is necessary to arrive at the conditions
for diffusion driven instability. A detailed derivation of the general
conditions for the diffusion driven instability can be found in
Murray[1989]. For discrete cases one can follow the same derivation as
in the case of continuous systems by considering solutions of the form
$\exp i(kj-\lambda t)$ [Murray 1989; Mu\~{n}uzuri {\it et al.}, 1995].
Here $k$ and $\lambda$ are considered to be independent of the position
$j$ $(j=1,2,\cdots,N)$. For Eq.  (3), the criteria for the diffusion
driven instability can be derived by finding the conditions for which
the steady states in the absence of diffusive coupling are linearly
stable and become unstable when the coupling is present.  One can
easily derive from a straightforward calculation that the following
conditions should be satisfied for the general reaction-diffusion
system of the form given by Eq. (\ref{eqn3all}):
\begin{eqnarray}
\label{cond:1}
f_x+g_y & < & 0,\nonumber \\
f_xg_y-f_yg_x & > &  0,\nonumber \\
f_xD_2-g_yD_1 & > & 0,\nonumber \\
(f_xD_2-g_yD_1)^2-4D_1D_2(f_xD_2-g_yD_1) & > & 0.
\end{eqnarray}
The critical wave number for the discrete system (3) can be obtained as
\begin{equation}
\label{cond:2}
\cos(k_c)=1-\displaystyle{f_xD_2-g_yD_1 \over 4D_1D_2}.
\end{equation}
Combining Eqs. (\ref{cond:1})-(\ref{cond:2}), one obtains[Mu\~nuzuri
{\it et al.}, 1995] the condition for the Turing
instability such that
\begin{equation}
\label{cond:3}
\displaystyle{f_xD_2-g_yD_1 \over 8D_1D_2} \le 1.
\end{equation}

We apply these conditions to the coupled oscillator system of the
present study. For this purpose, we fix the parameters for the
two-dimensional model (3) as $\{\beta, \sigma, \epsilon, m_0, m_1,
m_2\}=\{0.8,$ $ 0.92, 0.1, -0.5, 0.5, 0.5\}$ and verify that this
choice satisfies the conditions (\ref{cond:1}) to (\ref{cond:3}).  The
numerical simulations are carried out using an array of size $100\times
100$ and random initial conditions near the steady states are chosen
\begin{figure}[!ht]
\includegraphics[width=\linewidth]{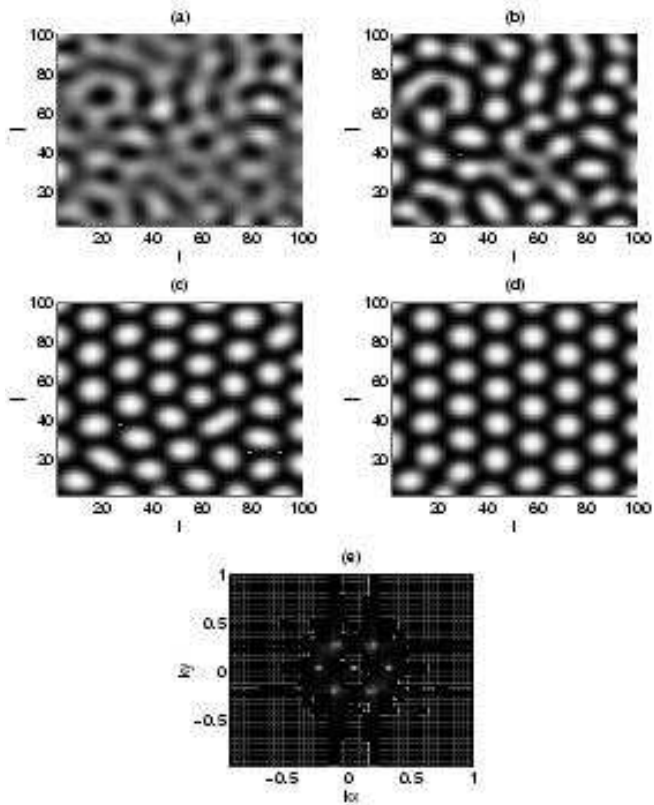}
\caption{
The spontaneous formation of Turing patterns in an array of $100\times 100$
oscillators for the parameters $\beta=0.8$, $\sigma=1$, $m_0=-0.5$, 
$m_1=0.5$, $m_2=0.5$, $\epsilon=0.1$, $F=0.0$, $\omega=0.75$, $D_1=1$ and 
$D_2=10$ in Eq. (3) at various time units (a) $T=50$, (b) $T=100$, (c) 
$T=500$ and (d) $T=2000$ (e) the same figure (d) in the Fourier space.
}
\label{fig5}
\end{figure}
for the $x$ and $y$ variables.  Figs. \ref{fig5}(a)-\ref{fig5}(d) show
how the diffusion driven instability leads to stable hexagonal pattern
(Fig. \ref{fig5}(d)) after passing through intermediate stages (Figs.
\ref{fig5}(a)-\ref{fig5}(c)).  Fig. \ref{fig5}(e) shows the two
dimensional Fourier spectrum of the hexagonal pattern (Fig.
\ref{fig5}(d)). Further, the spontaneously formed patterns are fairly
uniform hexagonal patterns having a penta-hepta defect pair. These
defects are inherent and very stable. We note that such Turing patterns
have already been observed in an array of coupled Chua's
circuits[Mu\~nuzuri {\it et al}, 1995]. However, our aim here is to
investigate the effect of external forcing on these patterns, which is
taken up in the next section.

%%%%%%%%%%%%%%%%%%%%%%%%%%%%%%%%%%%%%%%%%%%%%%%%%%%%%%%%%%%%%%%%%%%%%
\section{Spatio-temporal patterns in the presence of periodic external force}
\label{sec4}

The effect of external fields on a variety of dynamical systems has
been studied for a long time as driven systems are very common from a
practical point of view. For example, in a large number of dynamical
systems including the Duffing oscillator, van der Pol oscillator and
the presently studied MLC circuits, temporal forcing leads to a variety
of complex dynamical phenomena such as bifurcations, quasiperiodicity,
intermittency, chaos and so on[Guckenheimer \& Holmes, 1983; Hao
Bai-Lin, 1984; Lakshmanan \& Murali,1996].  Also the studies on the
effect of external fields in spatially extended systems have been
receiving considerably increasing interest in recent times [Pismen,
1987; Walgraef, 1988; 1996].  Particularly, with the recent advances in
identifying localized and oscillating structures and other
spatio-temporal patterns in driven nonlinear dissipative systems such
as granular media, driven Ginzburg-Landau equations and so on, it is of
considerable interest to study the effects of forcing on array of
coupled systems such as (3).  Motivated by the above, we investigate
the effect of external forcing in the propagation of wave front and
formation of Turing patterns in the coupled MLC circuits in one and two
dimensions.

\subsection{Effect of external forcing on the propagation of
wave fronts}
\label{sec4a}

In this subsection we study the effect of external forcing on the
propagation of wave fronts. For this purpose we consider an one
dimensional array of coupled MLC circuits with initial and boundary
conditions as discussed in Sec. \ref{sec3a}.  Now we perform the
numerical integration by the inclusion of external periodic force of
frequency $\omega=0.75$ in each cell of the array (see Eq. (1)).  By
varying the strength, $F$, of the external force we study the behaviour
of the propagating wave front in comparison with the force free case
($F=0$) as discussed in Sec. \ref{sec3a}. We find that in the
propagation region ($D>D_c$), the effect of forcing is just to
introduce temporal oscillations and the propagation continues without
\begin{figure}[!ht]
\includegraphics[width=\linewidth]{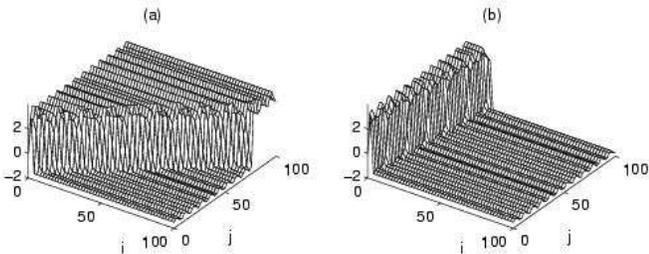}
\caption{
Propagation phenomena in the presence of forcing: (a) Propagation
of wavefront for $D=2.0$ and $F=0.6$ and (b) The partial propagation
observed for the $D=0.22$ and $F=0.6$.
}
\label{fig6}
\end{figure}

any disturbance (see Fig. \ref{fig6}(a)) as in the case $F=0$ (Fig.
\ref{fig2}(a)).  Of course this can be expected as the external force
is periodic in time.  However, interesting things happen in the
propagation failure region discussed in Sec. \ref{sec3a}. In this
region, beyond a certain critical strength of the external forcing, the
wavefront tries to move a little distance and then stops, leading to a
partial propagation. Fig. \ref{fig6}(b) shows such a partial
propagation observed for $F=0.6$ and $D=0.22$ (This may be compared
with Fig. \ref{fig2}(b)). The phenomenon can be explained by
considering the propagation failure mechanism discussed in Sec.
\ref{sec3b} in which one may look for a spatially stationary and
temporally oscillating wavefront. The initial wavefront tries to settle
in the nearby stationary state. However, the system will take a little
time and space to settle due the effect of forcing combined with the
transient behaviour of the system. Thus the inclusion of external
forcing in the propagation failure region can induce the wavefront to
achieve a partial propagation.

\subsection{Transition from hexagons to rhombs}
\label{sec4b}

It is well known that the defects are inherent in very many natural
pattern forming systems. In most of the pattern forming systems, the
observed patterns are not ideal. For example, the patterns are not of
perfect rolls or hexagons or rhombs. A commonly observed defect in such
systems is the so called penta-hepta defect (PHD) pair which is the
bound state of two dislocations [Tsimring, 1996]. Experiments on
spatially extended systems often show the occurrence of PHD in
spontaneously developed hexagonal patterns[Pantaloni \& Cerisier,
1983]. In the present case also, the existence of PHD pair can be
clearly seen from Fig. \ref{fig5}(d).  In such a situation, it is
important to study the effect of external periodic force in the coupled
arrays of MLC circuits.

Now we include a periodic force with frequency $\omega$ and amplitude
$F$ in each cell of the array and we numerically integrate
Eqs.(\ref{eqn3all}) using fourth order Runge-Kutta method with zero
flux at boundaries. By fixing the frequency of the the external
periodic force as $\omega=0.75$ and varying the amplitude ($F$) we
analyse the pattern which emerges spontaneously.  Interestingly for
$F=0.25$, the defects (PHD pair) which are present in the absence of
external force (Fig. \ref{fig5}(d)), gets removed resulting in the
\begin{figure}[!ht]
\includegraphics[width=0.9\linewidth]{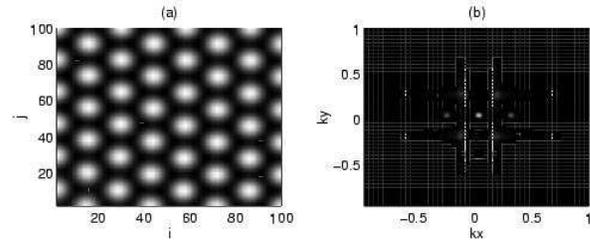}
\caption{
Figure showing the formation of rhombic structures in the 
presence of forcing. (a) Gray scale plot of the rhombic pattern
for $F=0.25$ and (b) The same figure in Fourier space.
}
\label{fig7}
\end{figure}

transition to a regular rhombic array.  Fig. \ref{fig7}(a) shows the
gray scale plot of the pattern observed for $F=0.25$ and the
corresponding plot in Fourier domain (Fig. \ref{fig7}(b)).  Thus, from
the above we infer that the inclusion of external periodic force can
cause a transition from hexagonal pattern to rhombic structures. We
note that P\'erez-Mu\~nuzuri {\it et al}.  have observed, in arrays of
coupled Chua's circuits, similar transition from hexagons to rhombs by
means of {\em sidewall} forcing [P\'erez-Mu\~nuzuri {\it et al.},
1995]. However in our case we apply forcing simultaneously to all the
cells, in order to mimic situations such as application in vertically
vibrated granular media[Umbanhowar {\it et al.}, 1996].

\subsection{Transition from hexagons to rolls}
\label{sec4c}

In addition to the transition from hexagons to rhombs by the influence
of external periodic force, there are also other possible effects due
to it. To realize them, we consider a different set of parametric
\begin{figure}[!ht]
\includegraphics[width=\linewidth]{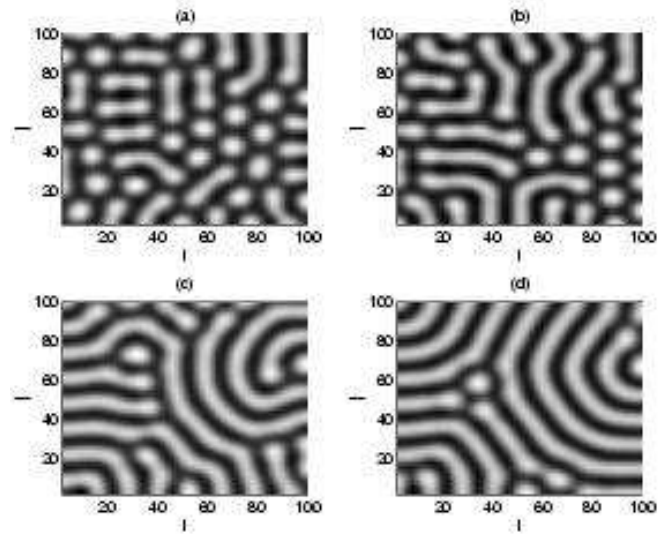}
\caption{
Figure showing the transition from hexagons to rolls for  $\{ \beta,$ 
$\sigma,$ $\epsilon,$ $m_0$, $m_1,$ $m_2\}$ $=$ $\{0.734722,$ $0.734722,$ 
$0.15,$ $-0.874,$ $-0.4715,$ $-0.4715\}$ with $D_1=1$ and $D_2=5$. (a) 
$F=0$, (b) $F=0.15$, (c) $F=0.35$ and (d) $F=0.45$.
}
\label{fig8}
\end{figure}
choice $\{\beta, \sigma, m_0, m_1, m_2\}$ $=$ $\{0.734722,$
$0.734722,$  $-0.874,$ $ -0.4715,$ $-0.4715\}$ with $\epsilon=0.15$,
$D_1=1.0$ and $D_2=5.0$.  For this choice the system shows hexagonal
patterns with defects including domains of small roll structures (Fig.
\ref{fig8}(a)).

Now when the external periodic force is included a transition in the
pattern from hexagonal structure to rolls starts appearing.  By fixing
the frequency of the external periodic force again at $\omega=0.75$, we
observed the actual transition from hexagons to rolls as we increase
the forcing amplitude ($F$).  Figs. \ref{fig8}(b)-\ref{fig8}(d) show
the gray scale plots for $F=0.15$, $0.35$ and $0.45$, respectively.
Obviously the transition is due to the existence of small roll
structures in the pattern for $F=0$ which nucleates the formation of
rolls in the presence of forcing.

\subsection{Breathing oscillations}
\label{sec4d}

In the above, we have shown that the inclusion of the external periodic
force can make a transition from one stationary pattern to another
stationary pattern like the transition from hexagons to rolls.  Besides
these, are there any time varying patterns? As mentioned above,
patterns such as localized and breathing oscillations have considerable
\begin{figure}
\includegraphics[width=0.9\linewidth]{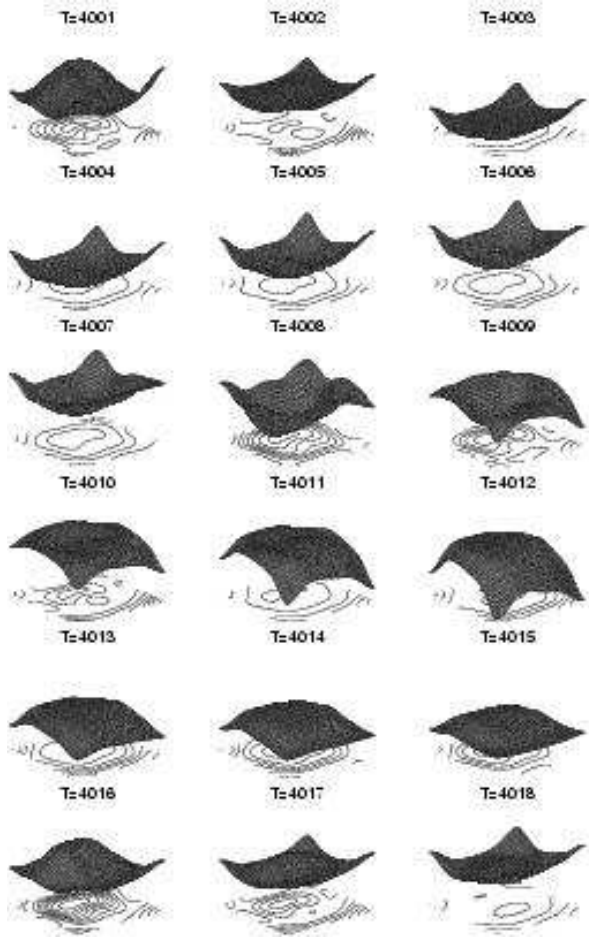}
\caption{
Snapshots showing the breathing oscillations for $F=0.05$ and $\{ \beta,$ 
$\sigma,$ $\epsilon,$ $m_0$, $m_1,$ $m_2\}$ $=$ $\{0.734722,$ $0.734722,$ 
$0.15,$ $-0.874,$ $-0.4715,$ $-0.4715\}$ with $D_1=2$ and $D_2=5$ for various
time units starting from $T=4001$.
}
\label{fig9}
\end{figure}
physical interest. In this regard, we considered the parameters $\{
\beta, \sigma, \epsilon, m_0, m_1, m_2\}$ $=$ $\{0.734722,$ $0.734722,$
$0.10,$ $-0.874,$ $ -0.4715,$ $-0.4715\}$ with $D_1=2.0$ and $D_2=5.0$
such that a regular hexagonal pattern is observed in the {\em absence}
of external periodic force.  From numerical simulations, we observed
that a space-time periodic oscillatory pattern (breathing motion) sets
in for a range of low values of $F$. Fig. \ref{fig9} shows the typical
snapshots of the oscillating pattern at various instants for the
specific choice $F=0.05$.  We have integrated over 10000 time units and
the figure corresponds to the region $T=4000 - 4018$. Typically we find
that the breathing pattern repeats itself approximately after a period
$T=15.0$ in the range of our integration. One may conclude that the
emergence of such breathing oscillations is due to the competition
between the Turing and Hopf modes in the presence of external periodic
force.

%%%%%%%%%%%%%%%%%%%%%%%%%%%%%%%%%%%%%%%%%%%%%%%%%%%%%%%%%%%%%%%%%%%%%
\section{Spatio-temporal chaos} 
\label{sec5}

Next we move on to a study of the spatio-temporal chaotic dynamics of
the array of coupled MLC circuits when individual cells are driven by
external periodic force. The motivation is that over a large domain of
($F$, $\omega$) values the individual MLC circuits typically exhibits
various bifurcations and transition to chaotic motion (see Appendix
A).  So one would like to know how the coupled array behaves
collectively in such a situation, for fixed values of the parameters.
For this purpose, we set the parameters at $\{\beta, \sigma, \epsilon,
m_0, m_1, $ $m_2, \omega\} = \{1.0, 1.015, 0, -1.02, -0.55, -0.55, 0.75
\}$. In our numerical simulations, we have mainly considered the one
dimensional array specified by Eq. (1) and assumed periodic boundary
conditions. The choice of periodic boundary conditions here makes the
calculation of Lyapunov exponents easier so as to understand the
spatio-temporal chaotic dynamics of (1) in a better way.

\subsection{Spatio-temporal regular and chaotic motion}
\label{sec5a}

Numerical simulations were performed by considering random initial
conditions using fourth order Runge-Kutta method for seven choices of
$F$ values. The coupling coefficient in Eq. (1) was chosen as $D=1.0$.
Out of these,  the first three lead to period-$T$, period-$2T$,
period-$4T$ oscillations, respectively and the remaining choices
\begin{figure}[!ht]
\includegraphics[width=0.9\linewidth]{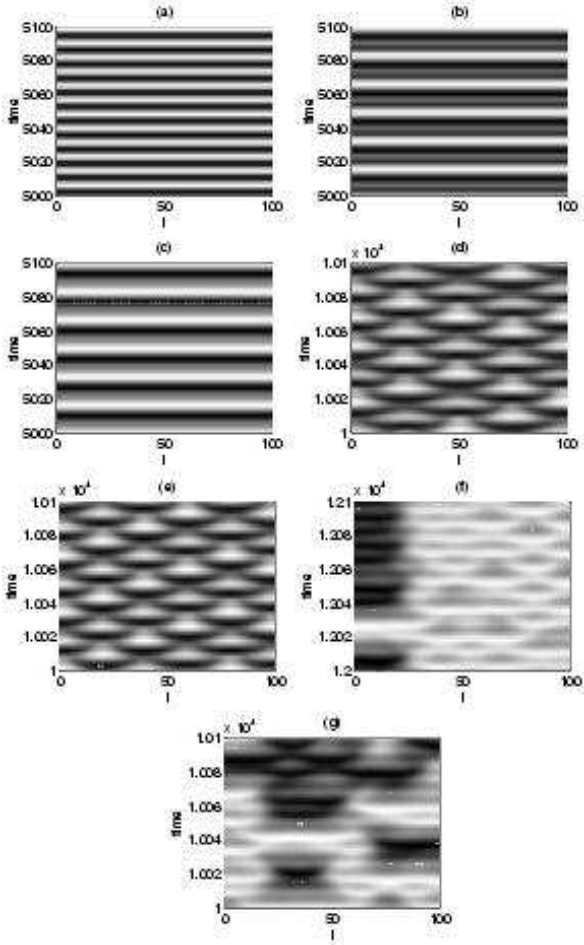}
\caption{
Spatio-temporal periodic oscillations in a grid of $100$ oscillators
for the external periodic forcing strength (a) $F=0.05$,  (b) $F=0.08$,
(c) $F=0.09$,  (d) $F=0.12$, (e) $F=0.13$, (f) $F=0.14$  and (g)
$F=0.15$ }
\label{fig10} 
\end{figure}
correspond to chaotic dynamics of the single MLC circuit. Figs.
\ref{fig10}(a)-\ref{fig10}(g) show the space-time plots for $F=0.05$,
$F=0.08$, $F=0.09$, $F=0.12$, $F=0.13$, $F=0.14$ and $F=0.15$,
respectively. From the Figs. \ref{fig10}(a)-\ref{fig10}(c), it can be
observed that for $F=0.05$, $0.08$ and $0.09$ the MLC array also
exhibits regular periodic behaviour with periods $T$, $2T$ and $4T$
respectively in time alone as in the case of the single MLC circuit.
However, for $F=0.12$ and $0.13$(Figs. \ref{fig10}(d) and
\ref{fig10}(e)) one obtains {\em space-time periodic} oscillations
eventhough each of the individual uncoupled MLC circuits for the same
parameters exhibits chaotic dynamics. We may say that a kind of
controlling of chaos occurs due to the coupling, though the coupling
strength is large here.  From the above analysis it can be seen that
the macroscopic system shows regular behaviour in spite of the fact
that the microscopic subsystems oscillate chaotically. Finally for
$F=0.15$ the coupled system shows spatio-temporal chaotic dynamics
(Fig. \ref{fig10}(g)) and this was confirmed by calculating the
Lyapunov exponents using the algorithm given by Wolf {\it et al}. [Wolf
{\it et al.}, 1985].  For example, we calculated the Lyapunov exponents
for $N=50$ coupled oscillators (For $N=100$, it takes enormous computer
time relatively which we could not afford at present) and we find the
highest three exponents have the values $\lambda_{\mbox{max}}=\lambda_0
=0.1001$, $\lambda_1=0.0776$, $\lambda_2=0.0092$ and the rest are
negative (see next subsection for further analysis). We also note that
for $F=0.14$, the system is in a transitional state as seen in Fig.
\ref{fig10}(f).

\subsection{Size dependence of the ST Chaos}
\label{sec5b}

Since the above study of spatio-temporal chaos involves a large number
of coupled chaotic oscillators it is of great interest to analyse the
size dependence of the dynamics of these systems. To start with, we
consider the case of 10 coupled oscillators with periodic boundary
conditions and numerically solve the system using fourth order
Runge-Kutta method with the other parameters chosen as in Sec.
\ref{sec5a}. The value of $F$ is chosen in the range $(0.12,0.15)$.  We
find that this set up shows different behaviour as compared to the 100
cell case. Actually the system gets synchronized to a chaotic orbit
rather than showing periodic behaviour as in the case of 100 cells
described above.

First we analyse the dynamics for $F=0.12$ by slowly increasing the
system size from $N=10$. We noted that for the system size, $N\le 42$
it gets synchronized to a chaotic orbit as mentioned above. We have
also calculated the corresponding Lyapunov spectrum and found that
there is one positive maximal Lyapunov exponent as long as $N\le 42$.
For example, for $N=42$, $\lambda_{\mbox{max}}=0.1162$, with the rest
\begin{figure}[!ht]
\includegraphics[width=0.9\linewidth]{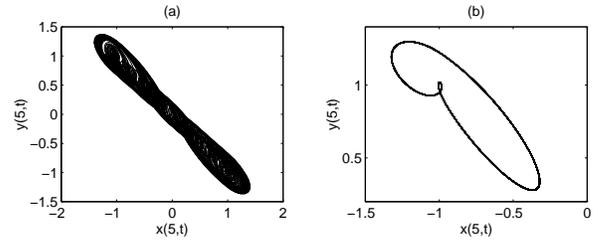}
\caption{
(a) The chaotic attractor at the $5^{\mbox{th}}$ cell for the synchronized
state $(N=42)$ and (b) the periodic orbit in the $5^{\mbox{th}}$ cell
for the controlled state $(N=43)$
}
\label{fig11}
\end{figure}
of the exponents being negative. Fig. \ref{fig11}(a) shows the dynamics
of the $5^{\mbox{th}}$ cell in the array and \ref{fig12}(a) depicts the
space-time plot of the synchronized spatio-temporal chaos. The system
shows entirely different behaviour when we increase the system size to
$N=43$.  As noted in the previous subsection, \ref{sec5a}, there occurs
kind of suppression of spatio-temporal chaos.  Fig. \ref{fig11}(b)
shows the resultant periodic orbit in the $5^{\mbox{th}}$ cell of the
array and Fig. \ref{fig12}(b) shows the space-time plot of the
spatio-temporal periodic behaviour of the array.  The maximal Lyapunov
exponent is found to be negative in this case $(\lambda_{\mbox{max}} =
-0.001474)$. We observed the same phenomenon for $F=0.13$ also.
\begin{figure}[!ht]
\includegraphics[width=0.9\linewidth]{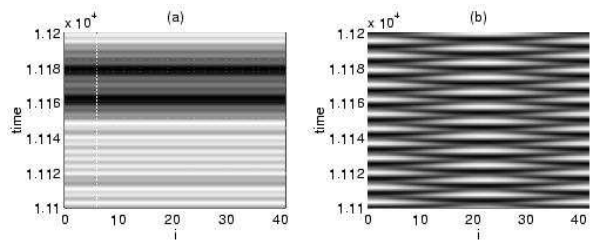}
\caption{
Space-time plot showing the size dependence of the dynamics in the
coupled chaotic oscillators. (a) $N=42$ and (b) $N=43$.
}
\label{fig12}
\end{figure}

Next we consider the case of $F=0.15$ in which the coupled system in
the previous subsection showed spatio-temporal chaos. From numerical
simulations, we again observed a synchronized motion for $N\le 31$ and
the corresponding Lyapunov spectrum shows one positive exponent only
with all other exponents being negative. But for $N>31$, the coupled
system shows spatio-temporal chaos. The Lyapunov spectrum in this case
(for $N>31$) possesses multiple positive exponents. For example, for
$N=43$, $(\lambda_{\mbox{max}}=\lambda_0 =0.0997$, $\lambda_1=0.0633$,
$\lambda_2=0.0038)$.

From the above analysis we see definite evidence that the
spatio-temporal chaotic dynamics in coupled MLC circuits depends very
much on the size of the system. A similar analysis can be performed to
investigate the nature of the spatio-temporal dynamics on the strength
of the coupling coefficient $D$. Preliminary analysis shows similar
phenomenon at critical values of $D$. Further work is in progress on
these lines.

%%%%%%%%%%%%%%%%%%%%%%%%%%%%%%%%%%%%%%%%%%%%%%%%%%%%%%%%%%%%%%%%%%%%%
\section{Conclusions}
\label{sec6}

      In this paper we have discussed various spatio\-temporal
behaviours of the coupled array of Murali-Lakshmanan-Chua circuits in
one and two dimensions.  In the  first part we have reported the
propagation phenomenon of the travelling wave fronts and failure
mechanism in the absence of external forcing. We have shown that the
propagation phenomenon is due to the loss of stability of the steady
states via subcritical bifurcation coupled with the presence of
necessary basin of attraction of the steady states. The effect of weak
coupling on the travelling wave phenomenon has also been studied.  In
the two dimensional array we studied the onset of Turing instability
leading to the spontaneous formation of hexagonal patterns.

Introduction of external periodic force has been shown to lead to a
partial propagation in the failure region of one dimensional array,
while in the two dimensional array a transition from hexagons to rhombs
take place. We have also showed that a transition from hexagons to
rolls can occur by the influence of external periodic force provided
domains of small roll structures already present in the absence of
external force. Space-time breathing oscillations as a result of the
competition between Hopf and Turing modes in the presence of external
periodic force have also been observed. We have also studied the
spatio-temporal chaotic dynamics and showed that the MLC array exhibits
both periodic and chaotic dynamics as in the case of single
oscillator.  In the chaotic regime the dynamics is also critically
dependent on the system size. For a large array, the collective
behaviour shows periodic oscillations both in space and time eventhough
the individual subsystems oscillate chaotically as the coupling brings
the macroscopic system into a regular motion.  On the other hand, below
a critical size of the system, it gets synchronized to a chaotic orbit.
The study of array of diffusively coupled driven nonlinear oscillators
is of intrinsic interest as they represent many natural phenomena such
as Faraday instability, patterns in granular media and so on. Of
particular interest will be to look for localized structures in these
systems. Whether such excitations exists in the present arrays of
oscillators is a question to be answered in the near future. Work is in
progress along these lines.

%%%%%%%%%%%%%%%%%%%%%%%%%%%%%%%%%%%%%%%%%%%%%%%%%%%%%%%%%%%%%%%%%%%%%
\acknowledgments P.M. acknowledges with gratitude the financial
support of Council of Scientific and Industrial Research in the form of
a Senior Research Fellowship. The work of M.L. has been supported by
the Department of Science and Technology, Government of India and
National Board for Higher Mathematics (Department of Atomic Energy) in
the form of research projects.
%%%%%%%%%%%%%%%%%%%%%%%%%%%%%%%%%%%%%%%%%%%%%%%%%%%%%%%%%%%%%%%%%%%%%
\appendix 
\section{Dynamics of the MLC circuit} 

\begin{figure}
\includegraphics[width=0.7\linewidth]{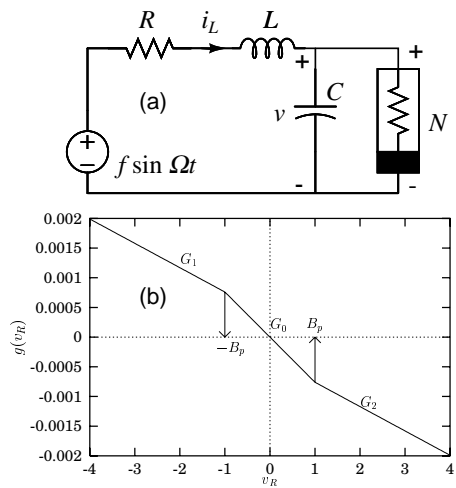}
\caption{ A single MLC circuit 
} 
\label{fig13}
\end{figure}

The Murali-Lakshmanan-Chua circuit, Fig. \ref{fig13}(a), is the
simplest second order dissipative nonautonomous circuit, consisting of
Chua's diode as the only nonlinear element[Murali {\it et al}., 1994a].
This circuit contains a capacitor $(C)$, an inductor $(L)$, a linear
resistor $(R)$, an external periodic forcing $(f\sin\Omega t)$ and a
Chua's diode.  By applying the Kirchoff's laws to this circuit, the
governing equations for the voltage $v$ across the capacitor $C$ and
the current $i_L$ through the inductor $L$ are represented by the
following set of two first order nonautonomous differential equations:

\begin{eqnarray} 
\label{app:eq1} 
C\frac{dv}{dt} & = & i_L-g(v),\nonumber \\ 
L\frac{di_L}{dt} & = & -Ri_L-R_si_L-v+f\sin\Omega t, 
\end{eqnarray} 

where $g(v)$ is a piecewise linear function
corresponding to the characteristic of the Chua's diode $(N)$ and is
given by 
\begin{equation} 
\label{app:eq2}
g(v_R)=\left\{
\begin{matrix}
\epsilon'+G_2v_R+(G_0-G_1), & v_R\geq B_p         \cr
\epsilon'+G_0v_R,           & -B_p\leq v_R\leq B_p \cr
\epsilon'+G_1v_R-(G_0-G_1), & v_R\leq -B_p
\end{matrix}
\right.
\end{equation} 

The piecewise nature of the characteristic curve of Chua's diode is
obvious from Eq. (\ref{app:eq2})[Murali {\it et al}., 1994a]. The
slopes of left, middle and right segments of the characteristic curve
are $G_1$, $G_0$ and $G_2$, respectively. $-B_p$ and $B_p$ are the
break points and $\epsilon'$ corresponds to the dc offset in the Chua's
diode.  Rescaling Eq.  (\ref{app:eq1}) as $v=xB_p$, $i_L=GyB_p$,
$G=1/R$, $\omega=\Omega C/G$ $t=\tau C/G$ and $\epsilon=\epsilon'/G$
and then redefining $\tau$ as $t$ the following set of normalized
equations are obtained:
\begin{eqnarray}
\label{app:eq3}
\dot x & = & y-h(x),\nonumber \\
\dot y & = & -\beta x -\sigma y +F\sin\omega t,
\end{eqnarray}
with
\begin{equation}
h(x)=\left\{
\begin{matrix}
     \epsilon+m_2x+(m_0-m_1), & x\geq x_2         \cr
     \epsilon+m_0x,           & x_1\leq x\leq x_2 \cr
     \epsilon+m_1x-(m_0-m_1), & x\leq x_1
\end{matrix}
\right., 
\end{equation}
where $\beta=(C/LG^2)$, $\sigma=(C/LG^2)(1+GR_s)$ and
$F=(f\beta/B_p)$.  Obviously $h(x)$ takes the form as in Eq.
(\ref{eqn2}) with $m_0=G_0/G$, $m_1=G_1/G$ and $m_2=G_2/G$. The
dynamics of Eq. (\ref{app:eq3}) depends on the parameters $\beta$,
$\sigma$, $m_0$, $m_1$, $m_2$, $\epsilon$, $\omega$ and $F$.

The rescaled parameters corresponding to the experimental observations
reported in  Murali {\it et al}.[1994a] correspond to $\beta=1.0$,
$\sigma=1.015$, $m_0=-1.02$, $m_1=-0.55$, $m_2-0.55$, $\epsilon=0$ and
$\omega=0.75$.  By varying $F$ one can observe the familiar period
doubling bifurcations leading to chaos and several periodic windows in
\begin{figure}[!ht]
\includegraphics[width=0.7\linewidth]{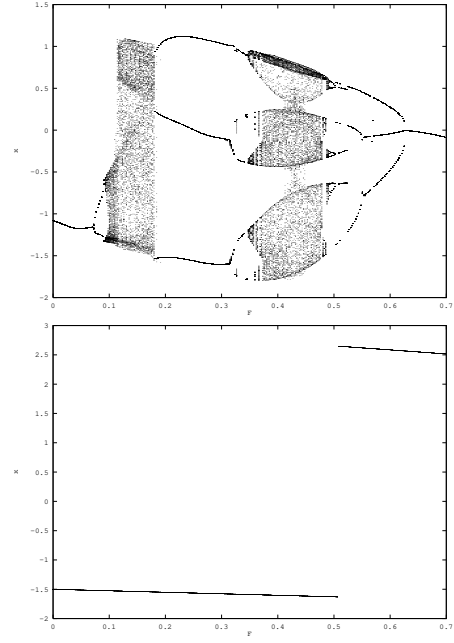}
\caption{
Bifurcation diagram in the $F$-$x$ plane of the MLC circuit for the 
parametric choice (a) $\{\beta, \sigma, m_0, m_1, m_2, \epsilon, \omega
\} $ $ = \{1.0, 1.015, -1.02, -0.55,-0.55, 0, 0.75\}$
and (b) $\{\beta, \sigma, m_0,$ $ m_1, m_2, \epsilon, \omega \} =
\{1.0, 1.0, -2.25,$ $ 1.5, 0.25, 0, 0.75\}$.
}
\label{fig14}
\end{figure}
the MLC circuit.  Fig. \ref{fig14}(a) shows the one parameter
bifurcation diagram in the $F$-$x$ plane [$F\in (0,0.7)$].  A summary
of the bifurcations that occur in this case for different $F$ values is
given in Table~\ref{table1}.  Further it is of great interest to consider the
parametric choice $\{m_0, m_1, m_2, \epsilon, \beta, \sigma, \omega\}
=  \{-2.25, 1.5, 0.25, 0.0,$ $1.0, 1.0, 0.75 \}$ which corresponds to
the function $h(x)$ having the form as shown in Fig. \ref{fig1}(c).
This choice of parameters provides the possibility of bistability
nature in the asymmetric case in the absence of periodic forcing. In
this case one can easily observe from numerical simulations that the
MLC circuit admits only limit cycles for $F\in (0,0.7)$. The
bifurcation diagram in the $F$-$x$ plane is depicted in Fig.
\ref{fig14}(b).
%%%%%%%%%%%%%%%%%%%%%%%%%%%%%%%%%%%%%%%%%%%%%%%%%%%%%%%%%%%%%%
\begin{table}
\caption{Summary of bifurcation phenomena of Eq.(A3) }
\begin{tabular}{ll}
\hline
amplitude $(F)$        & description of attractor \\ 
\hline
$0     <F\leq 0.071 $  & period-1 limit cycle \\ 
$0.071 <F\leq 0.089 $  & period-2 limit cycle \\ 
$0.089 <F\leq 0.093 $  & period-4 limit cycle \\ 
$0.093 <F\leq 0.19  $  & chaos \\ 
$0.19  <F\leq 0.3425$  & period-3 window \\ 
$0.3425<F\leq 0.499 $  & chaos \\ 
$0.499 <F\leq 0.625 $  & period-3 window \\ 
$0.625 <F           $  & period-1 boundary \\
\hline
\end{tabular}
\label{table1}
\end{table}
%%%%%%%%%%%%%%%%%%%%%%%%%%%%%%%%%%%%%%%%%%%%%%%%%%%%%%%%%%%%%%

In the main text, we also consider other choices of the parameters
depending on the phenomenon we are interested in.  The corresponding
values are given at the appropriate places in the text.
%%%%%%%%%%%%%%%%%%%%%%%%%%%%%%%%%%%%%%%%%%%%%%%%%%%%%%%%%%%%%%%%%%%%%%

\section{Steady states and Jacobian matrix of the system of 5 coupled
oscillators {(Eqs. \lowercase{(\ref{eqsta}))}}}

The steady states of Eq. (\ref{eqsta}) can be obtained explicitly as a
function of the diffusion coupling coefficient $D$ by introducing the
appropriate form of $h(x_i)$ from Eq. (\ref{eqn2}). Out of the
$3^5=243$ possible steady states we consider here a subset
$X_s=\{X_s^+, X_s^0, X_s^-\}$ and $X_p=\{X_p^+, X_p^0, X_p^-\}$
discussed in Sec. \ref{seclin}.

To obtain the set $X_s$ we proceed as follows. The components $x_{1s}$
and $x_{2s}$ are in the right extreme segment of the characteristic
curve, Fig. \ref{fig1}(c), while $x_{4s}$ and $x_{5s}$ are in the left
extreme segment. Therefore choosing $h(x_i)$ $=$ $m_2x+(m_0-m_1)$ for
$i=1,2$ and $h(x_i)$ $=$ $m_1x-(m_0-m_1)$ for $i=4,5$ (see Eq.
(\ref{eqn2})), we obtain
\begin{subequations}
\label{steady}
\begin{equation}
X^+_s  =  \{x_{1s}^+, y_{1s}^+, x_{2s}^+, y_{2s}^+, x_{3s}^+,
y_{3s}^+, x_{4s}^+, y_{4s}^+, x_{5s}^+, y_{5s}^+\}
\end{equation}
for $h(x_3)=m_2x+(m_0-m_1)$,
\begin{equation}
X^0_s  =  \{x_{1s}^0, y_{1s}^0, x_{2s}^0, y_{2s}^0, x_{3s}^0,
y_{3s}^0, x_{4s}^0, y_{4s}^0, x_{5s}^0, y_{5s}^0\}
\end{equation}
for $h(x_3)=m_0x$ and 
\begin{equation}
X^-_s  =  \{x_{1s}^-, y_{1s}^-, x_{2s}^-, y_{2s}^-, x_{3s}^-,
y_{3s}^-, x_{4s}^-, y_{4s}^-, x_{5s}^-, y_{5s}^-\}
\end{equation}
\end{subequations}
for $h(x_3)=m_1x-(m_0-m_1)$. Explicitly solving Eqs. (\ref{eqsta}) we 
can write
\begin{eqnarray}
x_{1s}^+ & = & \frac{3}{K}(16D^3+620D^2+550D+125), \nonumber \\
x_{2s}^+ & = & \frac{3}{K}(16D^3+500D^2+550D+125), \nonumber \\
x_{3s}^+ & = & \frac{3}{K}(16D^3+260D^2+400D+125), \nonumber \\
x_{4s}^+ & = & \frac{1.5}{K}(32D^3-200D^2-400D-125), \nonumber \\
x_{5s}^+ & = & \frac{1.5}{K}(32D^3-560D^2-550D-125),\nonumber \\
y_i^+ & = & -x_i^+, \,\,\,\,\,\, i=1,2,\ldots 5,\nonumber
\end{eqnarray}
\begin{eqnarray}
x_{1s}^- & = & -\frac{3}{L}(32D^4-920D^3-3200D^2-2750D-625), \nonumber \\
x_{2s}^- & = & -\frac{3}{L}(32D^4-560D^3-2000D^2-2000D-625), \nonumber \\
x_{3s}^- & = & -\frac{1.5}{L}(64D^4+320D^3+1700D^2+2000D+625), \nonumber \\
\end{eqnarray}
\begin{eqnarray}
x_{4s}^- & = & -\frac{1.5}{L}(64D^4+1280D^3+3500D^2+2750D+625), \nonumber \\
x_{5s}^- & = & -\frac{1.5}{L}(64D^4+1760D^3+3800D^2+2750D+625), \nonumber \\
y_i^- & = & -x_i^-,\;\;\;\;\;\; i=1,2,\ldots 5,\nonumber
\end{eqnarray}
\begin{eqnarray}
x_{1s}^0 & = & \frac{15}{M}(64D^3+44D^2-50D-25), \nonumber \\
x_{2s}^0 & = & \frac{15}{M}(48D^3+44D^2-30D-25), \nonumber \\
x_{3s}^0 & = & \frac{30D}{M}(8D^2+12D+5), \nonumber
\end{eqnarray}
\label{eqnb2}
\begin{eqnarray}
x_{4s}^0 & = & -\frac{7.5}{M}(32D^3-32D^2-40D-25), \nonumber \\
x_{5s}^0 & = & -\frac{7.5}{M}(64D^3+32D^2-50D-25),\nonumber \\
y_i^0 &= & -x_i^0, \,\,\,\,\,\, i=1,2,\ldots 5.
\end{eqnarray}
with
\[ K = 112D^3+620D^2+550D+125. \]
\[ L = 256D^4+1880D^3+3800D^2+2750D+625. \]
and
\[ M = 64D^4+320D^3+140D^2-250D-125. \]

The stability determining eigenvalues are the roots of the
characteristic equations of the Jacobian matrix for the linearised
version of Eq. (\ref{eqsta}):
\begin{widetext}
%{\small
\begin{equation}
\label{jacob}
J(X_s)=
\begin{pmatrix}
-h_{x_1}-D  & D & 0 & 0 & 0 & 1 & 0 & 0 & 0 & 0 \cr
D & -h_{x_2}-2D & D & 0 & 0 & 0 & 1 & 0 & 0 & 0 \cr
0 & D & -h_{x_3}-2D & D & 0 & 0 & 0 & 1 & 0 & 0 \cr
0 & 0 & D & -h_{x_4}-2D & D & 0 & 0 & 0 & 1 & 0 \cr
0 & 0 & 0  & D & -h_{x_5}-D & 0 & 0 & 0 & 0 & 1 \cr
-1 & 0 & 0 & 0  & 0 & -1 & 0 & 0 & 0 & 0     \cr
0 & -1 & 0 & 0  & 0 & 0 & -1 & 0 & 0 & 0     \cr
0 & 0 & -1 & 0  & 0 & 0 & 0 & -1 & 0 & 0     \cr
0 & 0 & 0  & -1 & 0 & 0 & 0 & 0 & -1 & 0     \cr
0 & 0 & 0  & 0 & -1 & 0 & 0 & 0 & 0 & -1      
\end{pmatrix}.
\end{equation}
%}
\end{widetext}
where $h_x$ is the derivative of $h(x)$ with respect to $x$ evaluated at
the steady states. 

Particularly we can show that for the trivial steady states
$X_p^+=(3.0$, $-3.0$, $3.0$, $-3.0$, $3.0$, $ -3.0$, $3.0$, $-3.0$, $3.0$, 
$-3.0)$, and
$X_p^-=(-1.5$, $1.5$,$-1.5$, $1.5$, $-1.5$, $1.5$, $-1.5$, $1.5$, $-1.5$, $1.5)$ are
stable and  $X_p^0 = (0$, $0$, $0$, $0$, $0$, $0$, $0$, $0$, $0$, $0)$ is unstable.  For
example considering $X_P^+$, the Jacobian (\ref{jacob}) takes the form

\begin{widetext}
%{\small
\begin{equation}
\label{jacob1}
J(X_P^+)=
\begin{pmatrix}
-1.5-D  & D & 0 & 0 & 0 & 1 & 0 & 0 & 0 & 0 \cr
D & -1.5-2D & D & 0 & 0 & 0 & 1 & 0 & 0 & 0 \cr
0 & D & -1.5-2D & D & 0 & 0 & 0 & 1 & 0 & 0 \cr
0 & 0 & D & -1.5-2D & D & 0 & 0 & 0 & 1 & 0 \cr
0 & 0 & 0  & D & -1.5-D & 0 & 0 & 0 & 0 & 1 \cr
-1 & 0 & 0 & 0  & 0 & -1 & 0 & 0 & 0 & 0     \cr
0 & -1 & 0 & 0  & 0 & 0 & -1 & 0 & 0 & 0     \cr
0 & 0 & -1 & 0  & 0 & 0 & 0 & -1 & 0 & 0     \cr
0 & 0 & 0  & -1 & 0 & 0 & 0 & 0 & -1 & 0     \cr
0 & 0 & 0  & 0 & -1 & 0 & 0 & 0 & 0 & -1      
\end{pmatrix}.
\end{equation}
%}
\begin{table*}[!hb]
\caption{
Stable steady states  $X_s^+$ and $X_s^-$ for $0\le D\le 0.4703$.  Only
$x_{is}, i=1,2\ldots 5$ are given. The values of $y$'s are obtained from the
relation $y_{is}=-x_{is}$ (One may note that the second initial conditions
$\{x_1(0),$ $x_2(0),$ $x_3(0),$ $x_4(0),$ $x_5(0) \}$ $=$ $\{2.7990,$ $2.1709,$
$1.7803,$ $-1.4433,$ $-1.4923\}$ and $y_i(0)=-x_i(0)$, $i=1,2,..5,$ chosen in
Sec. \ref{secnum}\ref{caseb} is closer to $X_s^+$ than $X_s^-$. Similarly the
first of the initial conditions $\{x_1(0),$ $x_2(0),$ $x_3(0),$ $x_4(0),$
$x_5(0) \}$ $=$ $\{3.0,$ $3.0,$ $-1.5,$ $-1.5,$ $-1.5\}$ and $y_i(0)=-x_i(0)$,
$i=1,2\ldots 5,$ is nearer to $X_s^-$ than $X_s^+$.)
}
\begin{tabular}{lllllll|lllllll}
\hline
&     $D$  & $x_{1s}$ & $x_{2s}$ &  $x_{3s}$ &  $x_{4s}$ &  $x_{5s}$ &
&     $D$  & $x_{1s}$ & $x_{2s}$ &  $x_{3s}$ &  $x_{4s}$ &  $x_{5s}$ \\ 
\hline
$X_s^+$ &   0.0000 &   3.0000 &   3.0000 &   3.0000 &  -1.5000 &  -1.5000 &
$X_s^-$ &   0.0000 &   3.0000 &   3.0000 &  -1.5000 &  -1.5000 &  -1.5000 \\
&   0.0100 &   3.0000 &   2.9997 &   2.9647 &  -1.4823 &  -1.4999 &
&   0.0100 &   2.9997 &   2.9647 &  -1.4823 &  -1.4999 &  -1.5000 \\
&   0.0200 &   3.0000 &   2.9989 &   2.9308 &  -1.4651 &  -1.4997 &
&   0.0200 &   2.9989 &   2.9308 &  -1.4651 &  -1.4997 &  -1.5000 \\
&   0.0300 &   2.9999 &   2.9977 &   2.8981 &  -1.4485 &  -1.4994 &
&   0.0300 &   2.9976 &   2.8981 &  -1.4485 &  -1.4994 &  -1.5000 \\
&   0.0400 &   2.9999 &   2.9960 &   2.8666 &  -1.4323 &  -1.4989 &
&   0.0400 &   2.9959 &   2.8666 &  -1.4323 &  -1.4989 &  -1.5000 \\
&   0.0500 &   2.9998 &   2.9939 &   2.8362 &  -1.4166 &  -1.4984 &
&   0.0500 &   2.9937 &   2.8362 &  -1.4166 &  -1.4984 &  -1.5000 \\
&   \ldots &   \ldots &   \ldots &   \ldots &   \ldots &   \ldots &
&   \ldots &   \ldots &   \ldots &   \ldots &   \ldots &   \ldots \\
%&   0.4000 &   2.9592 &   2.8316 &   2.1775 &  -1.0467 &  -1.4375 \\
&   0.4100 &   2.9571 &   2.8262 &   2.1655 &  -1.0393 &  -1.4351 &
&   0.4100 &   2.7921 &   2.1584 &  -1.0411 &  -1.4423 &  -1.4919 \\
&   0.4200 &   2.9550 &   2.8209 &   2.1537 &  -1.0321 &  -1.4327 &
&   0.4200 &   2.7853 &   2.1462 &  -1.0340 &  -1.4403 &  -1.4914 \\
&   0.4300 &   2.9528 &   2.8156 &   2.1422 &  -1.0250 &  -1.4303 &
&   0.4300 &   2.7784 &   2.1342 &  -1.0270 &  -1.4383 &  -1.4909 \\
&   0.4400 &   2.9506 &   2.8102 &   2.1308 &  -1.0180 &  -1.4279 &
&   0.4400 &   2.7715 &   2.1223 &  -1.0202 &  -1.4363 &  -1.4905 \\
&   0.4500 &   2.9484 &   2.8049 &   2.1196 &  -1.0111 &  -1.4254 &
&   0.4500 &   2.7646 &   2.1107 &  -1.0134 &  -1.4343 &  -1.4900 \\
&   0.4600 &   2.9461 &   2.7996 &   2.1087 &  -1.0043 &  -1.4230 &
&   0.4600 &   2.7577 &   2.0993 &  -1.0068 &  -1.4322 &  -1.4895 \\
&   0.4700 &   2.9435 &   2.7934 &   2.0937 &  -1.0162 &  -1.4234 &
&   0.4700 &   2.7508 &   2.0880 &  -1.0002 &  -1.4302 &  -1.4890 \\
\hline
\end{tabular}
\label{table2}
\end{table*}
\end{widetext}

The eigenvalues of the above matrix (\ref{jacob1}) as a function of $D$
can be obtained by solving the corresponding characteristic equation.
We have verified by numerical evaluation that all the eigenvalues of
(\ref{jacob1}) have negative real parts for $D\ge 0$. Similarly all the
eigenvalues corresponding to the steady state $X_p^-$ have negative
real parts irrespective of the values of $D$ $(D\ge 0)$. We have also
verified that atleast one of the eigenvalues corresponding to the state
$X_p^0$ has positive real part for $D\ge 0$. Thus $X_p^\pm$ are stable,
while $X_p^0$ is unstable for all values of $D\ge 0$.

Similarly one can obtain the stability properties of the set of steady
states $X_s=\{X_s^+, X_s^0, X_s^- \}$, by substituting the appropriate
expressions from (\ref{eqnb2}) into (\ref{jacob}). Numerical
diagonalization of (\ref{jacob}) for these steady states show that
$X_s^0$ is unstable for all values of $D$, while $X_s^+$ and $X_s^-$
are stable for $0\le D\le 0.4703$ and unstable for $D>0.4703$ for the
chosen initial conditions. The numerical values of $X_s^\pm$ in the
stable region are given in Table \ref{table2} as a function of $D$.

%%%%%%%%%%%%%%%%%%%%%%%%%%%%%%%%%%%%%%%%%%%%%%%%%%%%%%%%%%%%%%%%%%%%%%
x
%%%%%%%%%%%%%%%%%%%%%%%%%%%%%%%%%%%%%%%%%%%%%%%%%%%%%%%%%%%%%%%%%%%%%
\end{document}